\documentclass[conference]{IEEEtran}

\usepackage{multirow}
\usepackage{amssymb}
\usepackage{amsmath}
\usepackage{graphicx}
\usepackage{color}
\usepackage{subcaption}
\usepackage{units}
\usepackage{flushend}
\usepackage{microtype}
\usepackage{paralist}
\DeclareMathOperator*{\rms}{RMS}

\begin{document}

\title{Remixing Music with Visual Conditioning}

\author{\IEEEauthorblockN{Li-Chia Yang}
\IEEEauthorblockA{Center for Music Technology\\
Georgia Institute of Technology\\
Email: lyang351@gatech.edu}
\and
\IEEEauthorblockN{Alexander Lerch}
\IEEEauthorblockA{Center for Music Technology\\
Georgia Institute of Technology\\
Email: alexander.lerch@gatech.edu}
}

\maketitle

\begin{abstract}

We propose a visually conditioned music remixing system by incorporating deep visual and audio models. The method is based on a state of the art audio-visual source separation model which performs music instrument source separation with video information. We modified the model to work with user-selected images instead of videos as visual input during inference to enable separation of audio-only content. Furthermore, we propose a remixing engine that generalizes the task of source separation into music remixing. The proposed method is able to achieve improved audio quality compared to remixing performed by the separate-and-add method with a state-of-the-art audio-visual source separation model.

\end{abstract}

%
\IEEEpeerreviewmaketitle

\section{Introduction}

\begin{figure*}
    \includegraphics[width=\linewidth]{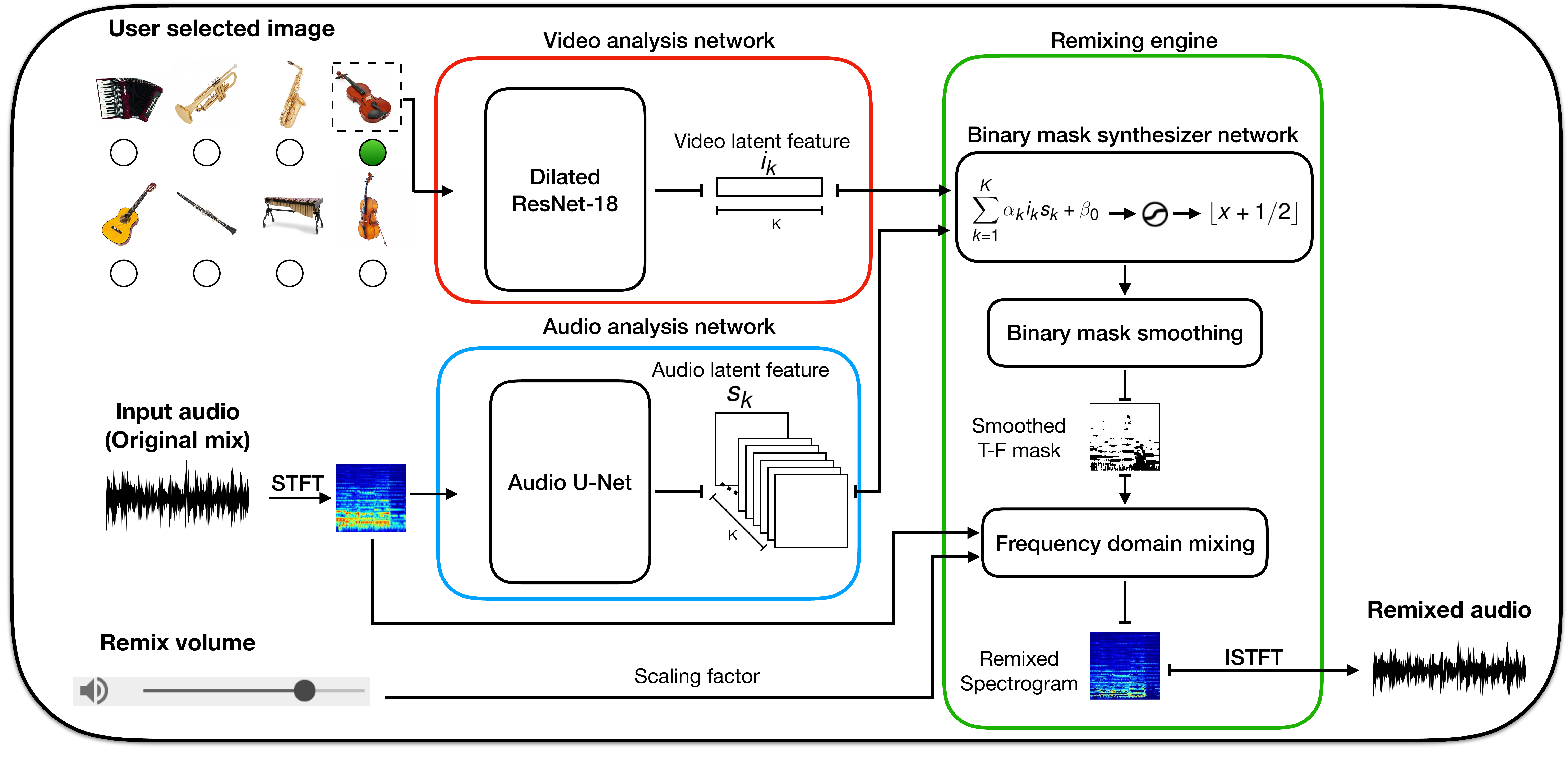}
    \caption{System diagram of the visually assisted remixer.}
    \label{fig:teaser}
\end{figure*}

In a music jam session with multiple instruments playing simultaneously, each musician is able to play synchronously with the others, focus on their and other parts, and continuously communicate not only through sound but also with visual cues. While a human processes and integrates both visual and auditory information automatically, most traditional systems for music analysis and processing focus on the acoustic signal exclusively and do not incorporate visual information.

The increasing availability of platforms with large computational resources enabled the development of approaches in both computer vision and audio that produce results with less error than humans \cite{sun2015deepid3} and can even be applied to tasks that can not be solved by human (e.g., a music recommendation system that analyzes the content of millions of audio files \cite{van2013deep}). There also has been some promising work on integrating visual and auditory patterns for classification tasks, e.g., \cite{lu2007survey, fu2011survey, lee2009unsupervised}. 

Following the success of these technologies, research proceeds to investigate applications that involve  integration of multiple sensory modalities such as the combination of audio and visual models. Examples for tasks where these approaches are popular are speech source separation \cite{Ephrat_2018} and music source separation \cite{Zhao_2018, zhao2019sound}. 

In the proposed work, we focus on the task of \textit{Music Remixing} which can be seen as a generalization of source separation (and suppression) approaches. In source separation systems, a weight of $1$ is assigned to the source of interest and weights of $0$ (mute) are assigned to the other sources. In music remixing, each source is assigned an individual adjustable gain, allowing the modification of the mix levels without complete separation of sources. While separation systems can theoretically yield the same remixing result by separating the sources and subsequently mixing them back together with different weights, we show below that remixing yields superior output quality for non-binary weights.

The proposed model is based on a state of the art audio-visual cross-modal music source separation model presented by Zhao et al.~\cite{Zhao_2018}, 
which uses solo instrument videos as training data and has a self-supervised training strategy that leverages the videos as ``visual cue" for the audio separation. 
Our model, however, performs inference with a user-assigned instrument image instead of a video recording as the visual cue.

To summarize, the main contributions of this work are
\begin{enumerate}
    \item   the introduction of the task \textit{Music Remixing} as a generalization of music source separation with improved audio quality over weighted mixing of completely separated sources, and
    \item   the demonstration of the feasibility and success of utilizing an instrument image (as opposed to a video) to allow for the  audio-only inference use case. 
\end{enumerate}

The remainder of the paper is structured as follows. The following Sect.~\ref{sec:relwork} presents a brief overview of relevant work. The proposed method is described in detail in Sect.~\ref{sec:method} and is evaluated in Sect.~\ref{sec:results}. We conclude in Sect.~\ref{sec:conclusion} with a brief summary and directions of future work.

\section{Related Work}\label{sec:relwork}
Source separation is the task of separating one or more audio signals representing the individual original sources from a weighted mixture of these signals \cite{bss, hyvarinen2000independent}. The most typical application areas of audio source separation are 
\begin{inparaenum}[(i)]
    \item   speech separation, with the goal of separating each speaker from a recording comprising multiple individuals speaking simultaneously \cite{bss_cocktail,cocktail}, and 
    \item   music source separation, which aims at separating the vocals and each musical instrument from a music recording \cite{duan2008unsupervised, simpson2015deep, luo2017deep, uhlich2017improving}.
\end{inparaenum}

Although several attempts have been made to apply rule-based algorithms defining constraints derived from characteristics of the target signals (compare, e.g., harmonic-percussive source separation \cite{fitzgerald2010harmonic}), the majority of approaches utilizes machine learning algorithms that leverage the independent structure within the mixture signal to adaptively filter out the set of possible components of the target signal. These approaches include, for example, Singular Value Decomposition \cite{zibulevsky2001blind}, Principal Component Analysis \cite{sansan2010blind}, and Non-negative Matrix Factorization \cite{ozerov2010multichannel, virtanen2007monaural}.

\begin{figure*}
\centering
  \includegraphics[width=\linewidth]{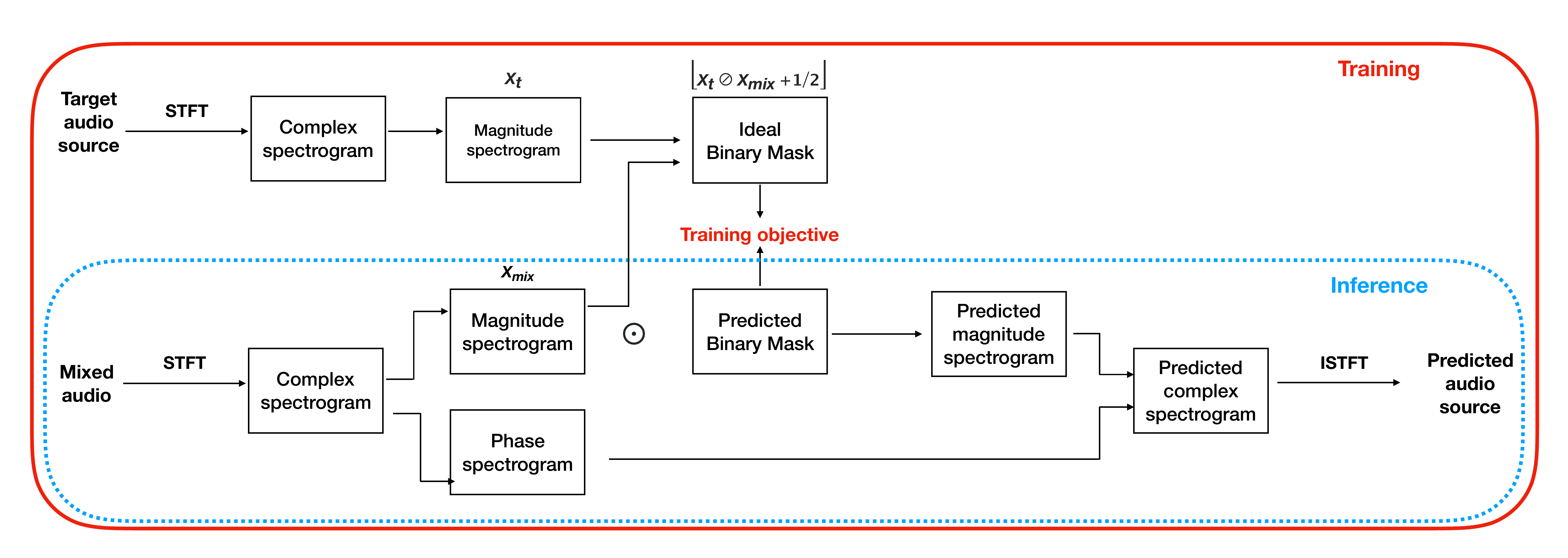}
\caption{Audio source separation with time-frequency masking.}
\label{fig: BSS_TF}
\end{figure*}

Modern deep learning systems have shown the capacity of modeling increasingly complex patterns in audio components. Network architectures such as Convolutional Neural Networks (CNNs) and Recurrent Neural Networks (RNNs) have been shown to significantly outperform traditional methods \cite{wang2018supervised, chandna2017monoaural}. 
Most modern audio source separation approaches estimate a time-frequency (T-F) mask for each instrument \cite{yilmaz2004blind} which is applied to the magnitude spectrogram of the input audio to separate the desired audio components as shown in Fig.~\ref{fig: BSS_TF}. Approaches for T-F masking vary mostly in the design of different training objectives such as signal approximation and mask approximation; the former minimizes the error between the separated audio signal and the target audio signal \cite{erdogan2015phase}, while the latter minimizes the error between the predicted mask and the target mask \cite{narayanan2013ideal,wang2013towards,xia2017using}. Zhao et al.\ compare two types of T-F masks, 
\begin{inparaenum}[(i)]
    \item   the \textit{Ratio Mask} (each pixel of the T-F mask $x \in [0,1]$), which has to be trained as a regression-like training task that is often hard to train, and
    \item   the \textit{Binary Mask} (each pixel of the T-F mask $x \in \left \{  0,1\right \}$) with a more classification-like training task.
\end{inparaenum}
The binary mask showed an overall better performance.

Several researchers proposed to combine audio and visual features to enrich the information that can be leveraged for either audio applications \cite{aytar2016soundnet, hershey2017cnn} or visual applications \cite{owens2016ambient}. 

Utilizing this cross-modality information, several systems have outperformed state-of-the-art single modality models in both speech separation and music source separation tasks \cite{Ephrat_2018, Owens_2018}.

Cross-modal learning can also solve the issue of data accessibility for supervised learning systems. It can allow to apply a self-supervised learning scheme \cite{sermanet2018time, zhong2017self} to label training data automatically. The visual information of unlabeled video data, for example, can be used as label to train a supervised learning model that separates audio \cite{Zhao_2018}. As manual annotation of video and audio data can be costly and time consuming, self-supervised learning provides a solution for scalable and efficient data collection.

Zhao et al.'s work titled ``The Sound of Pixels'' presents a music source separation system is able to provide the audio information by clicking an arbitrary pixel in the video frame of a music performance \cite{Zhao_2018}. The system is trained in a self-supervised framework along with the proposed ``MUSIC" dataset, which contains 536 untrimmed videos of solo instrument performance.
During the training phase, the system processes the visual information of solo videos with a video analysis network to extract the latent video feature, which is later being combined with the latent audio feature extracted by an audio analysis framework that takes the audio of mixture of multiple solo instruments as input. Finally, a synthesizer network leverages both latent features to generate the separated audio signal. As the latent video feature is designed to maintain the dimensions of the video frame, the system is able to provide the separated audio signal at the ``pixel level.''

\section{Method}\label{sec:method}

A re-implementation of Zhao et al.'s system serves as a baseline music source separation system as well as a starting point for the proposed remixing system. 

The proposed system features two main modifications as compared to the baseline system,
\begin{inparaenum}[(i)]
     \item   a modified inference mechanism using an instrument image as visual cue, and
     \item   a remixing engine that allows adjusting the gain of individual sources without having to separate the sources completely. 
\end{inparaenum}

\subsection{Implementation: Audio-Visual Source Separation}

The cross-modality architecture is comprised of a visual analysis model and an audio analysis model. The video analysis network is modified from Dilated ResNet-18 \cite{yu2017dilated}, with the last two layers removed and a $3\times 3$ convolution layer with $K$ output elements (corresponding to the size of the visual latent embedding) is added. As opposed to  Zhao et al., which extracts the visual feature as a tensor of $K \times X \times Y$ during the inference (a latent representation of length $K$ for each pixel in the image size of $X \times Y$), we perform spatial max pooling to the visual feature to collapse the embedding dimension to a vector of $K$ elements (compare the ``Video analysis network" in Fig.~\ref{fig:teaser}). This allows the network to keep only the most activated feature for the audio analysis model. The visual input representation computation follows Zhao et al., i.e., center cropping and down sampling to the shape of $224 \times 224$ and doubling the frame rate to 6 frames-per-second. A more detailed discussion is included in Sect.~\ref{sec:video}.

The audio analysis network has a modified Deep U-Net \cite{jansson2017singing} structure with 7 layers each for convolution and transposed convolution and skip connections in between. The Audio U-Net extracts a set of $K$ audio latent feature representations , matching the length of the visual latent feature, each with the audio input spectrogram dimensions (compare the ``Audio analysis network'' in Fig.~\ref{fig:teaser}). We design our experiment with a larger input audio representation than Zhao et al.\ as we use the audio sample rate of \unit[44.1]{kHz} (as opposed to \unit[16]{kHz}) to preserve more high frequency content. We then extract our input representation with an STFT window of size 1022 and a hop size of 512 frames on approximately 6 seconds of audio, resulting in an input matrix of dimensions $512\times 512$ (as opposed to $256 \times 256$).
After concatenating both the latent video and audio features, a synthesizer network estimates the T-F mask for the target audio source. 

The framework is trained in a self-supervised way by randomly selecting two solo instrument videos for the mix-and-separate framework. 

The binary T-F mask is chosen as our training target as Zhao et al. showed general superiority over the ration mask \cite{Zhao_2018}. The objective function of the training is the binary cross-entropy loss between the predicted and the ideal binary mask (compare Fig.~\ref{fig: BSS_TF}).
The network is trained for 85 epochs with a batch size of 16 and a learning rate of $1\mathrm{e}{-4}$ for the visual analysis network and a learning rate of $1\mathrm{e}{-3}$ for the both the audio analysis network and the mask synthesizer network.

\begin{table*}[t]
\begin{center}
\begin{tabular*}{\textwidth}{r|  @{\extracolsep{\fill}}c|cc|cc|cc}
\multicolumn{1}{c|}{}  & \begin{tabular}[c]{@{}c@{}}The Sound \\ of\\  Pixels\end{tabular} & \begin{tabular}[c]{@{}c@{}}VAreMixer\\ (video)\end{tabular} & \begin{tabular}[c]{@{}c@{}}VAreMixer\\ (image)\end{tabular} & \begin{tabular}[c]{@{}c@{}}VAreMixer\\ (wrong image)\end{tabular} & \begin{tabular}[c]{@{}c@{}}VAreMixer\\ (blank image)\end{tabular} & \begin{tabular}[c]{@{}c@{}}IBM\\ (best case)\end{tabular}   & \begin{tabular}[c]{@{}c@{}}RBM\\ (lower bound)\end{tabular}  \\ \hline
NSDR  & 8.87 & 7.83 & \textbf{9.05} & -5.88 & -17.71 & 13.69 & -3.73\\
SIR   & 15.02 & 14.11 & \textbf{15.77} & 4.10 & -0.15 & 22.15 & 1.45\\
SAR  & {12.28} & \textbf{10.63} & 10.46 & 2.66 & -11.73 & 15.10 & 1.05 
\end{tabular*}
\caption{Model performance evaluated with BSS Eval toolbox. IBM stands for ideal binary mask, RBM stands for random binary mask. No mask smoothing is applied.}
\label{table:result}
\end{center}
\end{table*}

\subsection{Inference with image as visual cue}\label{sec:video}
As mentioned above, the originally proposed system features a system called PixelPlayer, which allows the user to click on any pixel in the video frame to obtain the sound ``corresponding'' to the pixel's visual information \cite{Zhao_2018}. This PixelPlayer demonstrates that the audio-visual model is able to relate the audio pattern to granular visual features. However, the pixel level audio output is not entirely practical in the source separation scenario, where each instrument is mapped to exactly one target audio source. Furthermore, the pixel-to-sound concept has some fundamental inherent flaws; for example, it assumes that parts of the instrument which are visually masked do not produce sound. 
Practically, the network might also return inconsistent results when selecting different pixels of the same instrument.

Therefore, we base our design on the following three main considerations. 
First, each original sound source (instrument) should lead to only one separated target source.
Second, the audio-visual approach utilizing a self-supervised training strategy with solo videos as training data makes sense given the small size of traditional datasets.
Third, the requirement of a video recording during inference restricts possible use cases unnecessarily by neglecting audio-only input. Therefore, we design our approach to provide the separated target source given two inputs, the mixed audio recording and an image of the instrument to separate. This can be interpreted as a conditional audio source separation problem, however, it retains the advantage of self-supervised training.

The use of image as visual cue thus only targets the inference scenario while the training set-up remains unchanged. During the training phase, the visual analysis model is constructed by stacking an image network (Dilated ResNet-18) with the number of frames across the video, where the latent features of all frames are combined with temporal max-pooling operation. During the inference phase, this allows us to provide a single image as the only input frame to the image network and exclude the temporal max-pooling, resulting in the exact same output shape and dimension mapping. 

\subsection{Remixing engine}
The task of remixing can be interpreted as non-binary generalization of source separation. It is useful in scenarios where most components of the original mixture should remain but with different volume relations. While this could be achieved by completely separating all sources and subsequently weighting them and adding them together, our hypothesis is that a complete separation introduces more artifacts than applying the weight for different sources simultaneously. The proposed processing has two steps, mask smoothing and the frequency domain remixing itself.

\subsubsection{Binary mask smoothing}

Binary masks tend to introduce more artifacts than ratio masks during reconstruction due to the steep slopes or discontinuities in the time-frequency domain. For instance, discontinuities in the spectral axis can lead to an impulse response considerably longer than the Inverse Short-time Fourier Transform (ISTFT) window, which in turn leads to artifacts referred to as time-domain aliasing, while discontinuities in the time axis might lead to musical noise, ``ringing artifacts,'' or sudden volume changes.  These artifacts can be reduced by smoothing the binary mask with a low-pass filter. Here, we applied zero-phase filtering with an anticausal single-pole low-pass filter on the time axis 

\subsubsection{Frequency domain mixing}
As remixing does not require the complete separation of all source, the goal is to change the weighting of each individual source, given the mask the system has estimated for this source. The weighted superposition of the masked spectrograms allows to minimize discontinuity in the spectral domain.

The remixing of $N$ target separation signals is performed by setting scaling factors $s$ for each individual signal $i$ and the mixed signal $x_\mathrm{mix}$, respectively. The magnitude spectrogram of each individual signal $X_i$ is estimated by multiplying each individual smoothed mask $M_i$ with the input magnitude spectrorgram $X_\mathrm{mix}$. The final remixed magnitude spectrogram $X_\mathrm{remix}$ is : 

\begin{equation}
    X_\mathrm{mix}\cdot \left(1 + \sum\limits_{i=1}^{N}s_i \cdot M_i \right)
\label{eq:remix}
\end{equation}
Note that the range of each scaling factor $s_i$ could be  $-1\ldots \infty$, where the lower bound at $-1$ corresponds to muting the target source (In our interface demo, the upper bound is set to $+1$, corresponding to a \unit[+6]{dB} gain).

\section{Evaluation}\label{sec:results}
The evaluation of our model has two objectives,
\begin{inparaenum}[(i)]
    \item   the assessment of source separation quality to allow comparison with other models and to understand the influence of the image input on the output, and
    \item   the assessment of the remixing quality for different mixing weights.
\end{inparaenum}

Although the model is generally designed to separate one target instrument given one target image, it can be applied to an arbitrary number of sources. We perform the experiment with the setting of only two mixed sources to allow for better comparability with previous work and the more accessible and extensive assessment of the remixing quality.

\subsection{Data}
The data utilized is the ``MUSIC'' dataset \cite{Zhao_2018}, a dataset with musical recordings from YouTube selected by keyword query. We leverage developer tools such as youtube-dl \cite{youtube-dl} as YouTube data pipeline, and FFmpeg \cite{
FFmpeg} to handle multimedia file processing. The parsed dataset contains 493 untrimmed videos of musical solos spanning the 11 instrument categories accordion, acoustic guitar, cello, clarinet, erhu, flute, saxophone, trumpet, tuba, violin, and xylophone. We split the parsed dataset into a training set of 403 videos and an evaluation set of 91 videos.

Note that some of the originally listed videos of the dataset are no longer available. Thus, the resulting number of recordings (493) is necessarily smaller than that of the original dataset (536). 
For the instrument images, we perform a small scale data collection by gathering 10 images each for the 11 instrument categories.

A detailed list of our parsed dataset and image data can be found as supplementary material in our online repository.\footnote{https://github.com/RichardYang40148/VAreMixer}

\subsection{Source separation quality}

\begin{figure*}[t!]
    \centering
    \begin{subfigure}[t]{0.3\linewidth}
        \centering
        \includegraphics[height=2.5in]{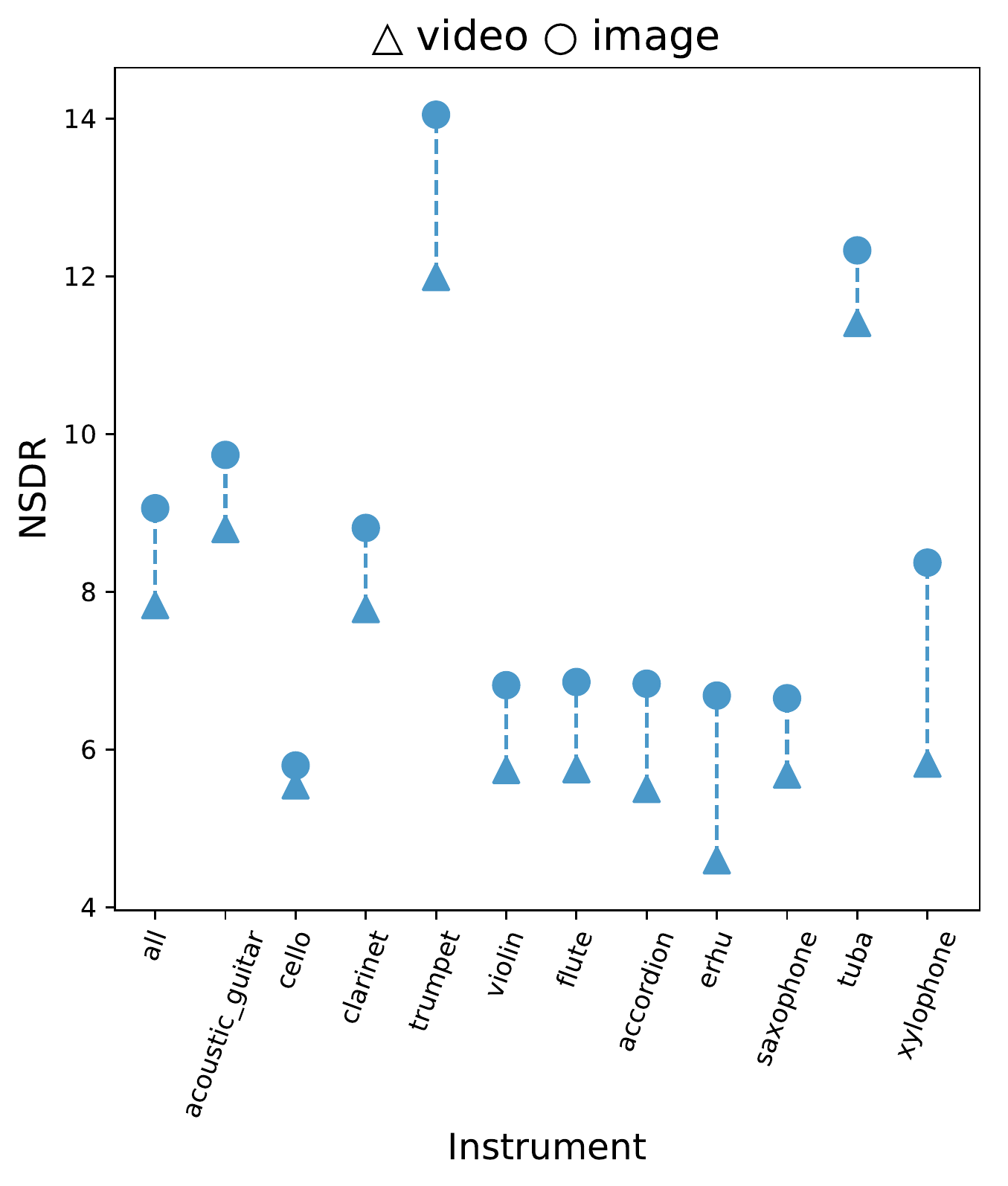}
        \caption{NSDR}
    \end{subfigure}
    ~
    \begin{subfigure}[t]{0.3\linewidth}
        \centering
        \includegraphics[height=2.5in]{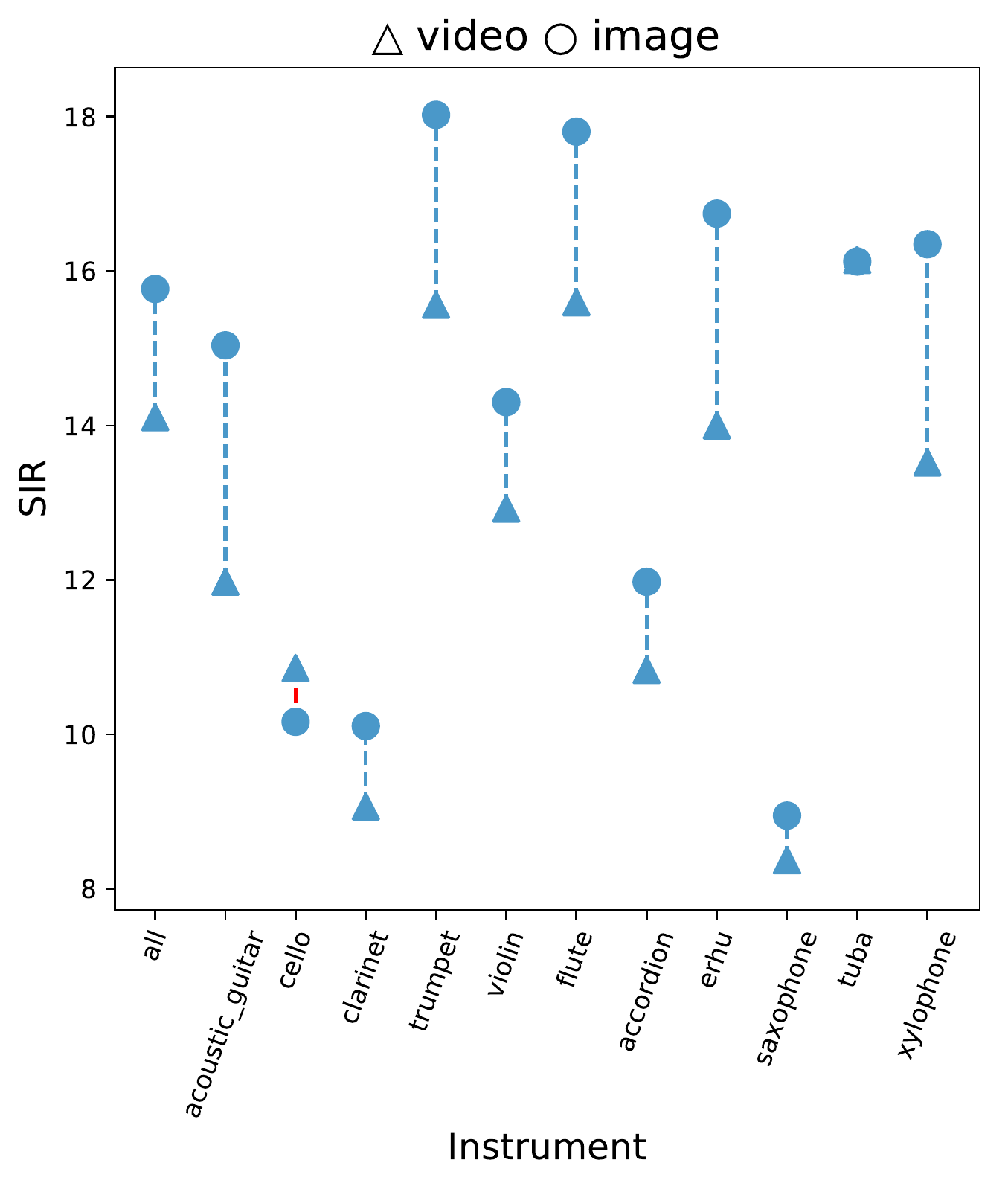}
        \caption{SIR}
    \end{subfigure}%
    ~
    \begin{subfigure}[t]{0.3\linewidth}
        \centering
        \includegraphics[height=2.5in]{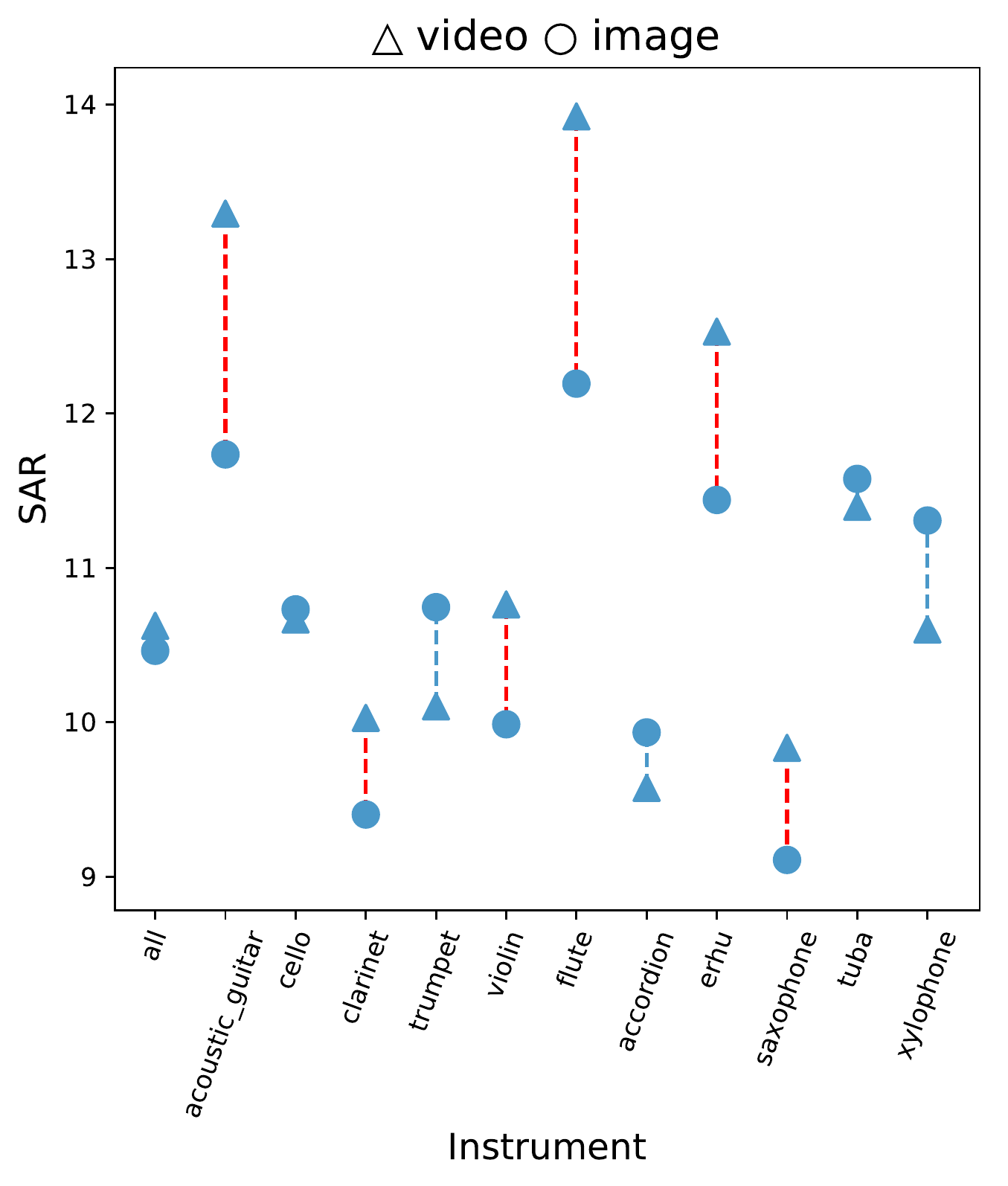}
        \caption{SAR}
    \end{subfigure}
    
    \caption{Instrument specific performance with video versus image as visual cue }
    \label{fig: instrument_eval}
\end{figure*}

In order to present comparable results to other work, the extreme case of remixing is evaluated first: source separation. The standard evaluation metrics for source separation are the three BSS Eval metrics \cite{bss}: Normalized Signal-to-Distortion Ratio (NSDR), Signal-to-Interference Ratio (SIR), and Signal-to-Artifact Ratio (SAR). Note that the NSDR is defined as the difference of SDRs of 
\begin{inparaenum}[(i)]
    \item   the separated signals compared to the ground truth signals, and 
    \item   the mixture signals compared to the ground truth signals. 
\end{inparaenum}
This provides a normalized SDR improvement measure independent of the complexity of the mixture signal.

The evaluation results are shown in Table~\ref{table:result}. We can make the following observations. 
First, note the IBM and RBM values; these provide an upper and lower bound, respectively. The Ideal Binary Mask (IBM) represents the separation results computed with a binary mask directly estimated from the ground truth and is, thus, a best case scenario. The results for the Random Binary Mask (RBM) are for a randomly generated mask, which simulates a system without inference capabilities. The range between those two is the range in which the results for every source separation system should lie. 

Second, comparing the results for our baseline VAreMixer (video) to the results originally reported in \cite{Zhao_2018}, we note that our baseline system is outperformed across all metrics. We speculate that the reason for this difference stems from two sources, 
\begin{inparaenum}[(i)]
    \item   the smaller size of our dataset (403:91 vs.\ 500:130) and the resulting different data split, and 
    \item   the increased input complexity due to the higher audio resolution (higher sample rate and longer STFT windows). 
\end{inparaenum}
We consider this reproduction difference to be only a minor issue, however, as we are mostly concerned with the relative changes of our system over the baseline. 

Third, it can be observed that our proposed image-based method VAreMixer (image) improves the result for the NSDR and the SIR over both, our baseline and the results previously posted. This increase can probably attributed to the higher image quality of our test image input compared to the video recordings. The SAR stays in the same range as the baseline system. A more in-depth analysis how the metrics compare between the baseline and the proposed system per instrument class is shown in Fig.~\ref{fig: instrument_eval}. It can be seen that while NSDR and SIR improve nearly across all instruments, about half of the instruments loose and half of the instruments gain in the SAR domain. However, we could not identify a specific common pattern across these instrument categories. Note that we neither claim nor expect superior performance over the original work \cite{Zhao_2018}. However, the results show that separation quality can improve or at least be on par despite using a less complex network for a more challenging inference task due to the missing video information.

Finally, we are interested in investigating how well the model understands the audio-visual correlation. We are doing this by
\begin{inparaenum}[(i)]
    \item   feeding a wrong image by randomly selecting images from instrument categories other than the target instrument as visual cue, and
    \item   feeding a blank image as visual cue.
\end{inparaenum}
The results confirm that the network behaves largely as expected: the wrong image decreases results for all metrics significantly and brings them close to the random mask boundary, while the blank image results in values considerably below that boundary as the output is mostly silent.

\begin{figure*}
\centering
\includegraphics[width=\linewidth]{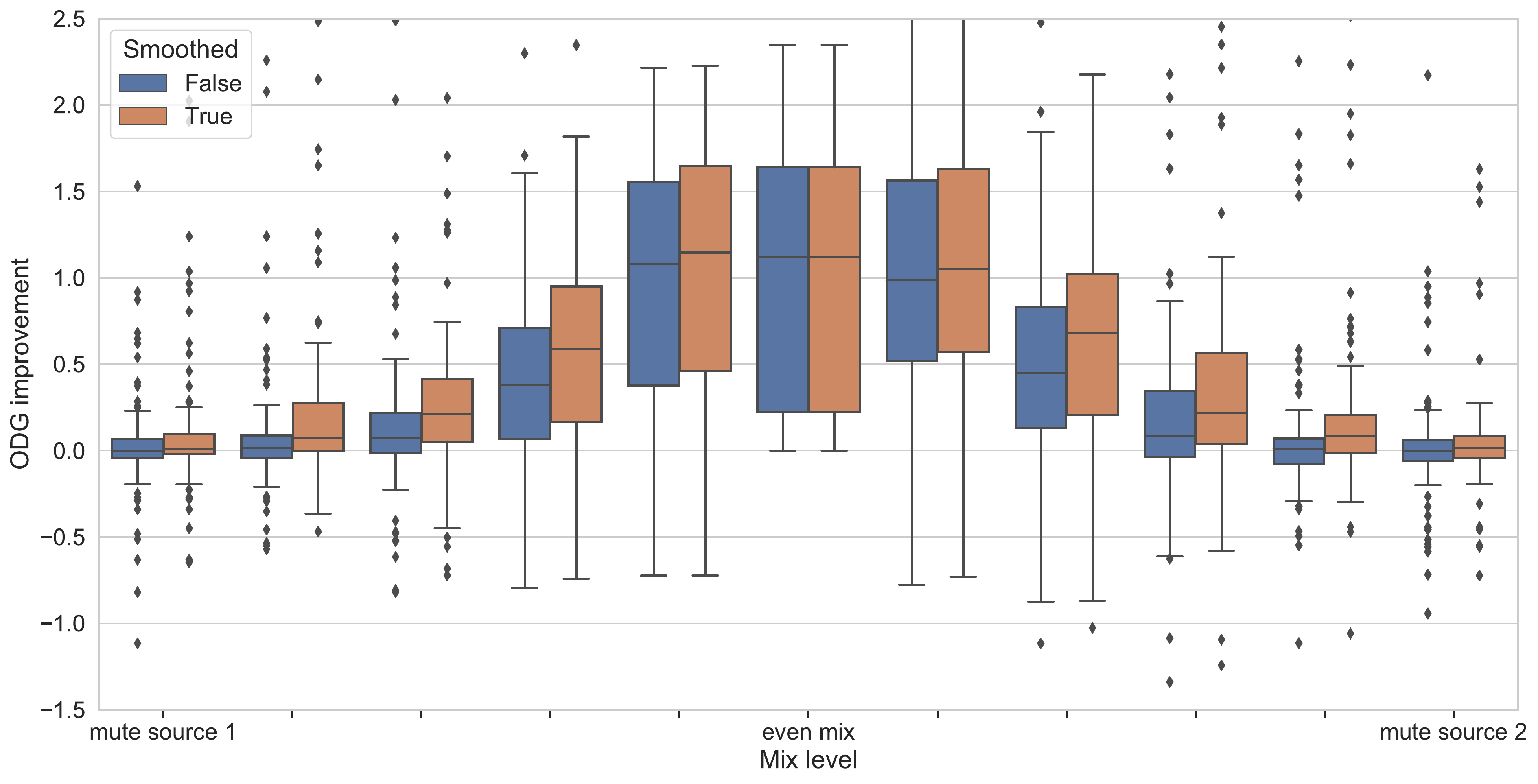}
\caption{Objective evaluation of perceptual quality. The plot shows the improvement of the ODG for different mixing levels for two sources (left and right) and with or without mask smoothing (blue and beige). The ODG is a measure on a scale of -4 to 0, An ODG improvement of 1 corresponds to the difference between neighboring ratings on the following scale: ``Imperceptible,'' ``Perceptible, but not annoying,'' ``Slightly annoying,'' ``Annoying,'' ``Very Annoying.''}
\label{fig: remix}
\end{figure*}

\subsection{Remixing quality}
To begin our quality evaluation, we first investigate the effect of binary mask smoothing. The baseline for comparison is the Cepstrally-Smoothed binary mask (CBM) as proposed by Stokes et al.~\cite{stokes2013reducing}. The experimental results of Stokes et al. show that the CBM, computed by the cepstral transform, can effectively improve the audio perceptual quality over binary T-F masking. To estimate the ideal smoothing parameter, we perform a hyperparameter search to both the method of Cepstrally-Smoothed binary mask (CBM) and our proposed zero-phase low-pass filtered binary mask (ZLBM) across our evaluation dataset.

The metric for comparison is computed as
\begin{equation}
    \Delta \mathrm{SNR} = 20\cdot \log_{10}\left(\frac{\rms(X_\mathrm{ref}-X_\mathrm{smooth})}{\rms(X_\mathrm{ref}-X_\mathrm{bin})}  \right).
\label{eq:dodg}
\end{equation}
The signal reconstructed by the ideal ratio mask is our reference signal $X_\mathrm{ref}$, the signal reconstructed by the binary T-F mask is $X_\mathrm{bin}$, and the signal reconstructed by the smoothed T-F mask as $X_\mathrm{smooth}$. The result shows $\unit[+0.065]{dB}$ gain on CBM and $\unit[+0.72]{dB}$ gain on ZLBM, indicating that compared to the Cepstrally-Smoothed binary mask, our proposed binary mask smoothing method yields a better reconstructed audio signal compared to the a Cepstrally-Smoothed binary mask.



Next, we proceed to the evaluation of the proposed frequency domain mixing strategy. While traditional source separation metrics cannot be applied as the sources are not separated, we can, however, compute three different signals for different mixing levels,
\begin{inparaenum}[(i)]
    \item   the ideal ground truth reference as the weighted superposition of the original sources,
    \item   the output of the proposed method, and
    \item   the output of the two-step process of first separating and then computing the weighted superposition.
\end{inparaenum}

The computed metric is the improvement of Objective Difference Grade (ODG) of the ``basic version'' of ITU-R recommendation BS.1387 \cite{kabal2002examination, recommendation19991387}, called Perceptual Evaluation of Audio Quality (PEAQ). Designed for the evaluation of audio coding quality with reference, the ODG utilizes a perceptual model to rate the quality of an input file compared to the unimpaired reference on a scale from -4 to 0. Figure~\ref{fig: remix} shows the absolute ODG improvement of the proposed method over the separate-and-add method for different mixing weights in the case of two sources mixed at equal level.

First, we can observe that the remixing engine provides the biggest quality improvement at an equal mixing level (volume both at 50\% for each instrument). Due to the spectral domain mixing strategy, our remixed signal is identical with the ground truth signal (($s_1=0$ and $s_2=0$ in Eq.~(\ref{eq:remix})). We can also see that the closer the relative mixing gain gets to the even mix, the higher the quality improvement is.

Second, at the extreme separation scenarios which keep only one source and mute the other, our magnitude spectrogram for signal reconstruction is calculated by subtracting the predicted magnitude spectrogram for the muted source from the magnitude spectrogram of the mix signal ($s_{i}$ for the muted source is set to $-1$ in Eq.~(\ref{eq:remix})). In this scenario, our spectral domain remixing strategy shows similar ODG scores to the separate-and-add approach as expected.

Third, we can confirm that the results computed with a smoothed mask (beige) give a slightly but consistently higher result across all mixing levels, reiterating our previous results.

Finally, there are some audio examples in which the ODG decreases for the frequency domain remixing cases. Some variability in the ODG results can be expected as PEAQ is simply an algorithmic model of listener test results that was developed to measure impairments due to coding artifacts. We therefore can speculate on the following reasons for measured quality decrease:
\begin{inparaenum}[(i)]
    \item   PEAQ model variability as the inference of the PEAQ model is imperfect, and 
    \item   PEAQ task specificity, as the focus on audio coding artifacts (that PEAQ is trained for) might not represent separation artifacts well in all cases.
\end{inparaenum}
Additionally, it is conceivable that there is uneven performance between two sources A and B: as the frequency domain strategy subtracts B from the mix to get A instead of using the A-mask directly, a bad mask estimation of source B might also impact the quality of source A. 

\subsection{Interface demo}

\begin{figure}
\centering
\includegraphics[width=\linewidth]{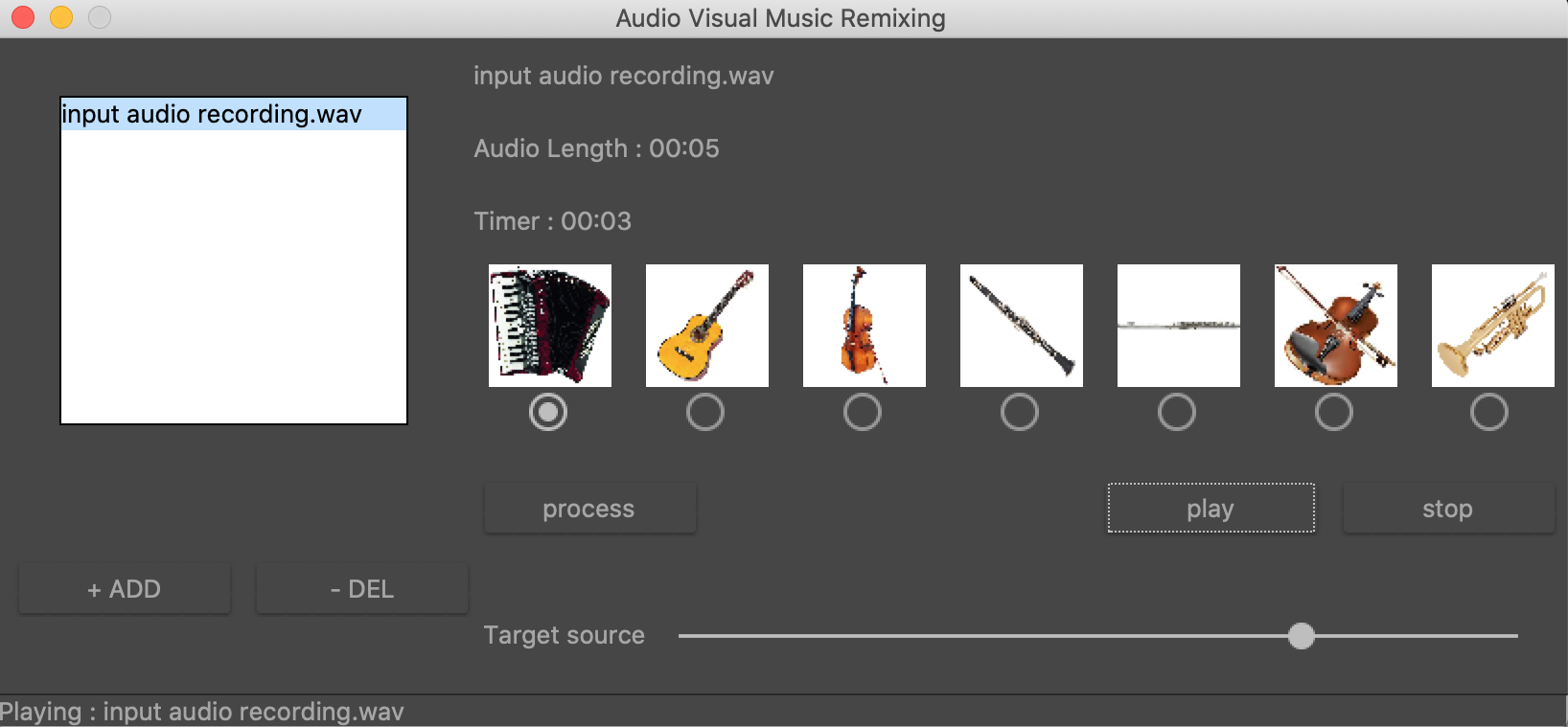}
\caption{User interface of the proposed remixing system.}
\label{fig: GUI}
\end{figure}

A demo of the presented system is available at our online repository.\footnotemark[1]

A screenshot of the interface as shown in Fig.~\ref{fig: GUI}. It prompts the user to load an audio recording and select one of the provided instrument images as target instrument. This image is then used as the visual cue for inference. The slider enables the user to adjust the volume of the target instrument in real-time.

\section{Conclusion}\label{sec:conclusion}

The project proposed a visually assisted music remixing system that integrates the knowledge of visual analysis and audio analysis networks to provide the ability to control the sound volume of individual instrument from a mixed sound source. Our experiment starts with the implementation of ``The Sound of Pixels,'' an audio-visual source separation model, and demonstrates improved performance with the method of using instrument video as training data, while using an image as visual cue during the model inference. 

In addition, by leveraging the difference between music remixing and source separation, we further exploit the spectral domain information from the original mix signal to reduce the overall artifact of reconstructed remix signal. As a result, we finalize our experiment with the development of VAreMixer.

There appears to be a significant performance drop for signals originating from non-acoustic instruments. This is expected in a data-driven approach where the training data does not include the samples to represent those instrument categories. Hence, the future work will focus on the extension of supported instrument categories and the adaptation to wider music genres. \newpage


\bibliographystyle{IEEEtran}
\bibliography{sample-base}

\end{document}